\documentclass[pra, aps, twocolumn, preprintnumbers,showpacs, floatfix,superscriptaddress]{revtex4}

\usepackage{latexsym}
\usepackage{mathrsfs}
\usepackage{amsmath}
\usepackage{amssymb}
\usepackage{amsfonts}
\usepackage{ifthen}
\usepackage{graphicx}
\usepackage{dcolumn}
\usepackage{bm}
\usepackage{color}
\usepackage{hyperref}
\begin{document}

\title{On the security of arbitrated quantum signature schemes}

\author{Qin Li}
\email[]{liqin805@163.com}
\affiliation{College of Information Engineering,
Xiangtan University, Xiangtan 411105, China}

\author{Chengqing Li}
\affiliation{College of Information Engineering,
Xiangtan University, Xiangtan 411105, China}
\affiliation{Department of Electronic and Information Engineering, The
Hong Kong Polytechnic University, Hong Kong}

\author{Zhonghua Wen}
\affiliation{College of Information Engineering,
Xiangtan University, Xiangtan 411105, China}

\author{Weizhong Zhao}
\affiliation{College of Information Engineering,
Xiangtan University, Xiangtan 411105, China}

\author{W. H. Chan}
\affiliation{ Department of Mathematics and Information Technology, The Hong Kong Institute of Education, Hong Kong}


\begin{abstract}
Due to potential capability of providing unconditional security, arbitrated quantum signature (AQS) schemes, whose implementation depends on the participation of a trusted third party, received intense attention in the past decade. Recently, some typical AQS schemes were cryptanalyzed and improved. In this paper, we analyze security property of some AQS schemes and show that all the previous AQS schemes, no matter original or improved, are still insecure in the sense that the messages and the corresponding signatures can be exchanged among different receivers, allowing the receivers to deny accepting the signature of an appointed message. Some further improvement methods on the AQS schemes are also discussed.


\end{abstract}

\maketitle

Digital signature,  as an electronic equivalent to hand-written signature in online transactions, is a very important cryptographic primitive and has many different uses. For instance, it can be used to authenticate the identity of the originator, ensure data integrity, and provide non-repudiation service. At present, classical (digital) signature has been widely used in electronic commerce and other related fields. Unfortunately, most existing classical signature schemes whose security depends on the difficulty of solving some hard mathematical problems were threatened by quantum computation \cite{Shor:AQC:FOCS94}. Therefore, researchers turn to investigate its quantum counterpart with the hope that quantum signature can become an alternative to classical signature and provide unconditional security.

Generally, a quantum signature scheme is believed to be unconditionally secure if the following two basic requirements are satisfied even though powerful quantum cheating strategies exist and unlimited computing resources are available: 1) the attacker (or the malicious receiver) cannot forge the signature; 2) disavowal of the signatory and the receiver is impossible. In 2002, unconditionally secure quantum signature was proved to be impossible by Barnum \emph{et al}. \cite{Barnum:AQM:FOCS02}. Even the result is disappointing, Zeng and Keitel proposed an arbitrated quantum signature (AQS) scheme with the aid of a trusted third party named arbitrator \cite{zeng:AQS:PRA02}. Afterwards, Li \emph{et al}. found that the arbitrator is unnecessary to entangle with the other two participants in the AQS scheme presented in Ref. \cite{zeng:AQS:PRA02} and thus the three-particle entangled GHZ states used in the scheme can be
replaced with two-particle entangled Bell states \cite{Qin:Sign:PRA09}. In addition, the preparation and distribution of Bell states are much easier to be implemented than that of GHZ states with the present-day technologies. So, Li \emph{et al.} proposed a more efficient AQS scheme using Bell states \cite{Qin:Sign:PRA09}. Zou \emph{et al}. showed both the two schemes proposed in Ref. \cite{zeng:AQS:PRA02} and Ref. \cite{Qin:Sign:PRA09} are insecure since they could be repudiated by the receiver Bob and presented two AQS schemes claimed to fix the secure problem \cite{Zou:Sign:PRA10}. But Hwang \emph{et al}. pointed out in Ref. \cite{Hwang:Sign:upd11} that the arbitrator cannot solve the dispute between the signatory Alice and the receiver Bob when Bob claims a failure in the verification phase of the scheme proposed by Zou \emph{et al.} Besides, some other security problems of these typical AQS schemes were also been discovered \cite{Chong:Sign:upd11,Gao:Sign:PRA11,Choi:Sign:upd11,Sun:Sign:upd11}.

In this paper, we study security of all the above mentioned AQS schemes \cite{zeng:AQS:PRA02,Qin:Sign:PRA09,Zou:Sign:PRA10} and find that
a common problem existing in the AQS schemes: different receivers can exchange their signed messages and the corresponding signatures arbitrarily, and thus they can deny the acceptance of the signature of an appointed message. The reason why this security problem exist is also analyzed in detail and the two AQS schemes presented by Zou \emph{et al}. \cite{Zou:Sign:PRA10} are selected as examples to study. In addition, we also discussed some potential improvement methods for enhancing the security of AQS schemes.

The rest of the paper is organized as follows. Section~\ref{sec:AQS1} introduces the AQS scheme with entangled states given in Ref. \cite{Zou:Sign:PRA10} and analyzes its security. Section~\ref{sec:AQS2} deals with the AQS scheme without entangled states proposed by Zou \emph{et al.} in \cite{Zou:Sign:PRA10}. Some discussions for improving the security of AQS schemes are given in Sec.~\ref{sec:dis}.
The last section concludes the paper.

\section{Security analysis of the AQS scheme with entangled states}
\label{sec:AQS1}

In this section, we will briefly introduce the AQS scheme with entangled states proposed in Ref. \cite{Zou:Sign:PRA10}, and then present security analysis
on it.

\subsection{The AQS scheme with entangled states}

The AQS scheme with entangled states proposed by Zou \emph{et al}. in Ref. \cite{Zou:Sign:PRA10} involves three participants, namely signatory Alice, receiver Bob, and the arbitrator, and consists of three
phases: the initializing phase, the signing phase, and the verifying
phase, which are described as follows.

\vspace{6pt} \noindent{\em {A. The initializing phase}}\vspace{6pt}

Step $I1$: The arbitrator shares keys $K_A$ and $K_B$ with Alice and Bob, respectively,
through quantum key distribution protocols proposed in Refs. \cite{BB:QKD:CSSP84,Ekert:QKD:PRL91},
which have been proved to be unconditionally secure \cite{lc:QKD:Science99,SP:QKD:PRL00}.

Step $I2$: Alice generates $N$ Bell states
$|\psi\rangle=(|\psi_1\rangle,|\psi_2\rangle,$ $\cdots,|\psi_N\rangle)$
with
$|\psi_i\rangle=\frac{1}{\sqrt{2}}(|00\rangle_{AB}+|11\rangle_{AB})$,
where the subscripts $A$ and $B$ correspond to Alice and Bob,
respectively. Then she distributes one particle of each
Bell state to Bob employing a secure and
authenticated method \cite{Barnum:AQM:FOCS02,Curty:QA:PRA02}.\vspace{6pt}

\noindent{\em{B. The signing phase}}\vspace{6pt}

Step $S1$: Alice transforms the message $|P\rangle$ into $|P'\rangle=E_r(|P\rangle)$ according to a randomly chosen
number $r\in\{00,01,10,11\}^N$.

Step $S2$: Alice generates $|S_A\rangle=E_{K_A}(|P'\rangle)$.

Step $S3$: Alice combines each message state and the Bell state to obtain the
three-particle entangled state
\begin{equation}
\begin{array}{lll}
    |\phi_i\rangle &=& |p_i'\rangle\otimes|\psi_i\rangle\\
                   &=&{\frac{1}{2}}\{|\phi_{12}^+\rangle_A(\alpha_i|0\rangle_{B}+\beta_i|1\rangle_{B})\\

                  & &+|\phi_{12}^-\rangle_A(\alpha_i|0\rangle_{B}-\beta_i|1\rangle_{B})\\

                  & &+|\psi_{12}^+\rangle_A(\alpha_i|1\rangle_{B}+\beta_i|0\rangle_{B})\\

                  & &+|\psi_{12}^-\rangle_A(\alpha_i|1\rangle_{B}-\beta_i|0\rangle_{B})\},
    \end{array}
    \label{eq:1}
    \end{equation}
where $|\phi_{12}^+\rangle_A$, $|\phi_{12}^-\rangle_A$,
$|\psi_{12}^+\rangle_A$, and $|\psi_{12}^-\rangle_A$ represent the four
Bell states respectively \cite{KMWZ:Bell:PRL95}.

Step $S4$: Alice implements a Bell measurement on each $|\phi_i\rangle$ and obtains $M_A=(M_A^1,M_A^2,\cdots,M_A^N)$, where $M_A^i$ represents one of
the four Bell states.

Step $S5$: Alice transmits the signature $|S\rangle=(|P'\rangle, |S_A\rangle, $ $|M_A\rangle)$ to Bob.

\vspace{6pt}\noindent{\em C. The verifying phase}\vspace{6pt}

Step $V1$:  Bob encrypts $|P'\rangle$ and $|S_A\rangle$ using the key
$K_B$ and sends the resultant outcome
      $|Y_B\rangle=E_{K_B}(|P'\rangle,|S_A\rangle)$ to the arbitrator.

Step $V2$: The arbitrator decrypts $|Y_B\rangle$ with $K_B$ and gets
$|P'\rangle$ and $|S_A\rangle$. Then he encrypts $|P'\rangle$ with
$K_A$ and obtains $S_T$. If $|S_T\rangle=|S_A\rangle$,
the arbitrator sets the verification parameter $V=1$, otherwise sets
$V=0$.

Step $V3$: The arbitrator obtains $|P'\rangle$ from $|S_T\rangle$
and sends the encrypted results
$|Y_{T}\rangle=E_{K_B}(|P'\rangle, |S_A\rangle, r)$ to Bob.

Step $V4$: Bob decrypts $|Y_T\rangle$ and obtains $|P'\rangle$, $|S_A\rangle$, and $r$.
If $r=0$, Bob rejects the signature, otherwise Bob makes further verification.

Step $V5$: According to Alice's measurement outcomes $M_A$ and Eq.
(\ref{eq:1}), Bob obtains $|P_B'\rangle$ via teleportation. If $|P_B'\rangle\neq|P'\rangle$, Bob rejects the signature, else
informs Alice to publish $r$.

Step $V6$: Alice announces $r$ through the pubic board.

Step $V7$: Bob recovers $|P\rangle$ from $|P'\rangle$ according to $r$ and takes $(|S_A\rangle,r)$ as the final signature of the message $|P\rangle$.

\subsection{Security analysis}

Hwang \emph{et al}. presented the deniability dilemma in the above AQS scheme \cite{Hwang:Sign:upd11}. They found the arbitrator cannot solve the dispute if Bob claims $|P_B'\rangle\neq|P'\rangle$ in Step $V5$ since the following three cases may occur: 1) Bob told a lie; 2) Alice sent a incorrect information to Bob; and 3) Eve disturbed the communication. However, if Bob made such an allegation, the verification process cannot be completed and a new signature task should be started. So, here we show that the receiver Bob can repudiate the acceptance of a signature related to a given message after finishing the verification process successfully.

First let Alice sign the message $|P\rangle_B$ for Bob and the message $|P\rangle_C$ for Charlie. Actually, $|P\rangle_B$ is favorable to Charlie, and $|P\rangle_C$ is beneficial to Bob. Then Bob and Charlie can be shown to exchange their messages and the corresponding signatures by using the following method. In step $I2$, after Alice distributes particles of Bell states to Bob and Charlie, Bob and Charlie exchange the particles they get. Similarly, after step $S5$, Bob sends the qubit string $|S\rangle_B=(|P'\rangle_B,|S_A\rangle_B,|M_A\rangle_B)$ to Charlie and Charlie returns
$|S\rangle_C=(|P'\rangle_C,|S_A\rangle_C,|M_A\rangle_C)$ to Bob. Then Bob can verify the validity of the signature $|S_A\rangle_C$ for the message $|P\rangle_C$ with the help of the arbitrator, and Charlie can check whether $|S_A\rangle_B$ is the signature of $|P\rangle_B$ with the aid of the arbitrator. Obviously, if Alice's signatures are valid, Bob and Charlie can finish the verification processes successfully. After that, Bob gets Alice's signature for the message $|P\rangle_C$ and Charlie obtains Alice's signature related to the message $|P\rangle_B$. Therefore, even if there are disagreements between Alice and Bob or between Alice and Charlie afterwards, Bob still can deny accepting the signature $|S_A\rangle_B$ of the message $|P\rangle_B$, and Charlie also can disavow the acceptance of the signature $|S_A\rangle_C$ related to the message $|P\rangle_C$. Furthermore, the arbitrator is not able to settle the dispute since they passed the verification processes.

\section{Security analysis of the AQS scheme without entangled states}
\label{sec:AQS2}

This section reviews the AQS scheme without entangled states proposed by Zou \emph{et al.} in Ref. \cite{Zou:Sign:PRA10}, and then analyzes the security of the scheme.

\subsection{The AQS scheme without entangled states}

The AQS scheme without entangled states also involves three participants, namely signatory Alice, receiver Bob, and the arbitrator, and consists of the following three phases.

\vspace{6pt} \noindent{\em {A. The initializing phase}}\vspace{6pt}

Step $I1$: The arbitrator shares keys $K_A$ and $K_B$ with Alice and Bob, respectively. In addition, Alice and Bob shares the key $K_{AB}$.

\vspace{6pt} \noindent{\em{B. The signing phase}}\vspace{6pt}

Step $S1$: Alice chooses a random number $r\in\{0,1\}^{2N}$ and computes $|P'\rangle=E_r(|P\rangle)$ and $|R_{AB}\rangle=M_{K_{AB}}(|P'\rangle)$.

Step $S2$: Alice generates $|S_A\rangle=E_{K_A}(|P'\rangle)$.

Step $S3$: Alice generates the signature $|S\rangle=E_{K_{AB}}(|P'\rangle,$ $|R_{AB}\rangle,|S_A\rangle)$ and transmits it to Bob.

\vspace{6pt}\noindent{\em C. The verifying phase}\vspace{6pt}

Step $V1$:  Bob obtains $|P'\rangle$, $|R_{AB}\rangle$, and $|S_A\rangle$ by decrypting $|S\rangle$ with the key
$K_{AB}$. Then he generates $|Y_B\rangle=E_{K_{B}}(|P'\rangle,|S_A\rangle)$ and sends it to the arbitrator.

Step $V2$: The arbitrator decrypts $|Y_B\rangle$ with $K_{B}$ and gets
$|P'\rangle$ and $|S_A\rangle$.

Step $V3$: The arbitrator obtains $|P_T'\rangle$ from $|S_A\rangle$
and compares it with $|P'\rangle$. If $|P_T'\rangle=|P'\rangle$, he sets the verification parameter $V_T=1$, else sets $V_T=0$.
The arbitrator announces the value of $V_T$ via the public board. If $V_T=1$, he reproduces $Y_B$ and resends it to Bob.

Step $V4$: If $V_T=0$, Bob rejects the signature, otherwise Bob decrypts $|Y_B\rangle$ and obtains $|P'\rangle$ and $|S_A\rangle$. Then he computes $|P_B'\rangle=M_{K_{AB}}^{-1}(|R_{AB}\rangle)$ and compares it with $|P'\rangle$. If $|P_B'\rangle=|P'\rangle$, he sets the verification parameter $V_B=1$, else sets $V_B=0$.
Bob announces the value of $V_B$ via the public board.

Step $V5$: If $V_B=0$, Alice and the arbitrator abort the scheme, otherwise Alice announces $r$ through the public board.

Step $V6$: Bob recovers $|P\rangle$ from $|P'\rangle$ by $r$ and takes $(|S_A\rangle,r)$ as Alice's final signature of the message $|P\rangle$.

\subsection{Security analysis}

In this subsection, we show that the arbitrator also cannot solve the disagreements between signatory and receiver for
the AQS scheme without entangled states if the following case happens.

Suppose Alice intend to sign the message $|P\rangle$ for Bob. Afterwards, Bob finds the message $|P\rangle$ is useless or unfavorable to him but beneficial to Charlie. Then by doing the following steps, Charlie can get the signature for $|P\rangle$ without being detected by Alice.
\begin{itemize}
\item First, when Bob receives $|S\rangle=E_{K_{AB}}(|P'\rangle,$ $|R_{AB}\rangle,|S_A\rangle)$ related to the message $|P\rangle$ from Alice after step $S3$, he decrypts it with the key $K_{AB}$ and obtains $|P'\rangle$, $|R_{AB}\rangle$, and $|S_A\rangle$. In addition, Bob gets another version of $|P'\rangle$ by decrypting $R_{AB}$ with the key $K_{AB}$.
\item Second, Bob transmits two versions of $|P'\rangle$ and $|S_A\rangle$ to Charlie through an authenticated channel.
\item Third, after Charlie has received what Bob sent, he encrypts $|P'\rangle$ and $|S_A\rangle$ with the key $K_C$ shared with the arbitrator to obtain $|Y_C\rangle=E_{K_C}(|P'\rangle,|S_A\rangle)$.
\item At last, the encrypted result $|Y_C\rangle$ is sent to the arbitrator.
\end{itemize}
Apparently, Charlie can implement the verification procedure like a honest receiver and get the signature of $|P\rangle$ if it is a valid one made by Alice. Furthermore, the arbitrator and Alice cannot discover the fact. Therefore, if there are disputes between Alice and Bob, Bob can deny that he has accepted the signature of the message $|P\rangle$, and Charlie can claim the signature of $|P\rangle$ does come from Alice if disagreements between Alice and Charlie exist.

\section{Possible Enhancements}
\label{sec:dis}

In this section, we first analyze two reasons why AQS schemes are easy to suffer deniability dilemma problem, and then propose the
corresponding improve methods.

One reason is that the signatory Alice cannot identify the real receiver. In other words, there is no relationship between the signed message and the real receiver. Therefore, different receivers can exchange their messages and the corresponding signatures
arbitrarily and thus repudiate accepting signatures for appointed messages. Another reason is that when participants announce random numbers or values of verification parameters, the identities of them and the announcement time are not published together. So, the arbitrator cannot distinguish which opened information is related to a specified message during a certain period.

According to the above analysis, we can take the following three measures to enhance the security of AQS schemes.
\begin{itemize}
\item First, the signatory Alice's signature not only includes the message, but also the identity of the receiver. Although the property of receivers' deniability is not always necessary in a signature scheme, it is quite useful in some special circumstances. For instance, suppose Alice sign a contract with Bob for a thousand dollars goods. If Bob can deny that he has accepted the contract with the help of another receiver Charlie and ask Alice to do the same thing again, it is quite unfair for Alice.
\item Second, when participants are required to announce random numbers or values of verification parameters, their identities and the announcement time should be also attached. So the arbitrator and the signatory can distinguish when the verification of signatures related to appointed messages is implemented and who participate in the verification process.
\item Third, before the signatory Alice start a signature procedure, she can tell the arbitrator who will be the receiver at first.
\end{itemize}

\section{Conclusion}

In this paper, we have shown two typical AQS schemes still suffer the security problem, namely receivers can deny any signature for an appointed message after the AQS procedures have been completed successfully. That is because a signed message is unrelated to a receiver, which allows different receivers to interchange their messages and the corresponding signatures arbitrarily. In addition, some countermeasures also have been presented to fix such a security problem. Whether these countermeasures can overcome all the security problems discovered in Refs. \cite{Chong:Sign:upd11,Gao:Sign:PRA11,Choi:Sign:upd11,Sun:Sign:upd11} and how to design an AQS scheme which can withstand existing or potential attacks deserve further research.

\section*{Acknowledgement}

This work is supported by Natural Science Foundation of China (Grant Nos. 61070232, 61100216, and 61105052), Scientific Research Fund of Hunan Provincial Education Department (Grant No. 11B124), and Science Fund of Xiangtan University (Grant No. 2011XZX16).

\bibliographystyle{apsrev4-1}
\bibliography{AQSS}

\end{document}